\documentstyle[version2,preprint,aps]{revtex}
\begin{document}
\draft
\begin{title}
{Freezing of the quantum Hall liquid at $\nu = $ 1/7 and 1/9}
\end{title}
\author{Rodney Price, Xuejun Zhu, and P.\ M.\ Platzman}
\begin{instit}
AT\&T Bell Laboratories, Murray Hill, New Jersey 07974
\end{instit}
\author{Steven G.\ Louie}
\begin{instit}
Department of Physics, University of California, Berkeley, and, \\
Materials Sciences Division, Lawrence Berkeley Laboratory,
California 94720
\end{instit}

\begin{abstract}
We compare the free energy computed from the ground state energy and
low-lying excitations of the 2-D Wigner solid and the fractional
quantum Hall liquid, at magnetic filling factors $\nu = 1/7$ and 1/9.
We find that the Wigner solid melts into the fractional quantum Hall
liquid at roughly the same temperature as that of some recent
luminescence experiments, while it remains a solid at the lower
temperatures characteristic of the transport experiments.  We propose
this melting as a consistent interpretation of both sets of
experiments.
\end{abstract}

\pacs{ }

\narrowtext
\newpage

The phase boundary between the fractional quantum Hall liquid, or
Laughlin liquid (LL), and the Wigner solid (WS) has been a subject of
active research for some time now.  Recent transport experiments
\cite{willett88,jiang90,goldman90} have shown clear evidence of an
insulating state that first appears just above $\nu = 1/5$,
disappears as $\nu$ approaches 1/5, then reappears strongly at all
smaller $\nu$.
This insulating phase has generally been interpreted
to be a manifestation
of a pinned Wigner solid, although recently an alternative theory
\cite{Kivelson} has been put forward where disorder plays a central
role in driving the phase transitions.

The ground state in a two-dimensional electron system
at filling factor $\nu$ = 1/7 and 1/9 remains unclear. Experimentally,
there is a weak signature in the transport data at $\nu$ = 1/7
\cite{jiang90,goldman88}
at somewhat elevated temperatures in the best samples available
today. But it disappears upon lowering the temperature to that
typical of magneto-transport experiments,
$\sim$ 20 - 100 mK. At $\nu$ = 1/9, there has been no report
of any transport anomaly.
On the other hand, magneto-optical luminescence experiments
\cite{buhmann90,buhmann91}, done at somewhat higher temperatures of
400 - 600 mK, show features in the luminescence spectrum at $\nu =
1/7$ and 1/9 that are similar to those at
$\nu$ = 1/3 and 1/5. The authors have
interpreted these features as due to the formation of the
quantum Hall liquid
at these filling factors. To explain the
discrepancy with the transport
studies, they suggest that the background resistivity, caused by magnetic-field
induced localization due to disorder, becomes so high at $\nu<1/5$ that the
quantum Hall states are unobservable in transport.

In this paper, we compare the free energies of the Wigner solid and
the Laughlin liquid at $\nu = $ 1/7 and 1/9, and we find that at
temperatures approximating those of the luminescence experiments the
Wigner solid will melt into a Laughlin liquid, but at the
temperatures of the typical
transport experiments, the Wigner solid remains
the stable state.  Our calculation is necessarily only
semi-quantitative, as disorder effects at these low filling factors
will have a significant effect on the ground-state energy of the
solid and the roton gap of the liquid.
Nonetheless, our results appear to
provide a consistent interpretation for both sets of experiments.

Previous calculations of the ground-state energy of the liquid and
solid \cite{makizotos83,lam84} have used variational wavefunctions
confined to the lowest Landau level.  Then the only relevant
dimensionless parameter is the magnetic filling factor $\nu$, and
it was found \cite{lam84}
that the melting
transition takes place at $\nu_c = 1/6.5$.  Physically, this high
magnetic field limit implies that as the magnetic field $B$ goes to
infinity the magnetic length $\ell$, where $\ell^2 = \hbar c/eB$,
goes to zero and the ion-disk radius $a = (\pi n)^{-1/2}$ goes to
zero, but the filling factor $\nu = 2\ell^2/a^2$ remains constant.
This works well at zero temperature, but at finite temperature, we
find ourselves comparing thermal energies of the order $k_BT$ to
Coulomb energies that go as $e^2/\ell \rightarrow \infty$, and we
must forgo the assumption that no Landau-level mixing occurs and
introduce another dimensionless parameter $r_s = a/a_B$, where $a_B =
\hbar^2 \epsilon/m^*e^2$ is the Bohr radius, $\epsilon$ is a
dielectric constant, and $m^*$ is an effective mass.  For the
electron-doped GaAs heterojunctions used in the experiments we use
the values $\epsilon = 12.8$ and $m^* = 0.068$.  Then a finite value
$B$ of the magnetic field requires a finite ion-disk radius $a$, and
we can reasonably compare thermal energies $k_BT$ to Coulomb energies
of the order $e^2/a$.

Price, Platzman, and He \cite{price93} have recently calculated the
ground-state energy of the Laughlin liquid at $\nu = $ 1/7 and 1/9 as
a function of $r_s$, using a variational wavefunction that included
Landau-level mixing.  Zhu and Louie \cite{zhu93} have also calculated
the ground-state energy of the Wigner solid at these filling factors
when $r_s = 2$ and 20, and they find that the ground-state energy of
the solid is lower than that of the liquid for all $r_s$.  Platzman
and Price \cite{platzman93} have also calculated the free energy of
both liquid and solid at $\nu = $ 1/3 and 1/5, and find a curious
reentrant freezing behavior in certain ranges of $r_s$ as the
temperature is raised from zero.  Here we will use the same method to
find the free energies at $\nu$ = 1/7 and 1/9, and we will provide an
estimated melting temperature as a function of $r_s$.

Because the temperatures of interest are so low, we need only know
the lowest-lying modes $\omega_k$ of both the solid and the liquid to
calculate the free energy.  Then the free energy is
\begin{equation}
F = E + T \sum_k \log(1-e^{-\omega_k/T}),
\label{free_energy}
\end{equation}
where $E$ is the ground-state energy and the $\omega_k$ are the
lowest lying excitations of either the liquid or the solid.  For the
Wigner solid these excitations are the lower-hybrid, essentially transverse
magnetophonons
$\omega_k^{WS}$.  We evaluated the free energy $F^{WS}$ of the solid
using the ground-state energy $E^{WS}$ from \cite{zhu93}, and the
harmonic magnetophonons $\omega_k^{WS}$ calculated in the same way as
{\it e.g.}, Bonsall and Maradudin \cite{bonsall77}, but with a strong
magnetic field.  The sum in
(\ref{free_energy}) for
the solid phase was evaluated by averaging over the Brillioun
zone by the method of Cunningham \cite{cunningham74}, so the result
is exact in the harmonic approximation.  It is useful, however, to
examine the form of the transverse
magnetophonons in the
long-wavelength limit
\begin{equation}
\omega_k^{WS} \approx 0.526\, \left(\frac{\nu}{r_s}
\frac{e^2/\epsilon}{2a_B}\right) (ka)^{3/2}.
\end{equation}
Substituting into (\ref{free_energy}) we find
\begin{equation}
F^{WS} \approx E^{WS} - 0.701\, \left(\frac{r_s}{\nu} \frac{2
a_B}{e^2/\epsilon} \right)^{4/3} T^{7/3},
\label{free_solid}
\end{equation}
and the free energy of the solid phase goes as $T^{7/3}$ at low
temperatures.

On the liquid side, since we are interested primarily in the
lowest-lying modes of the magnetoroton spectrum, we can approximate
the magnetoroton mode by
\begin{equation}
\omega_k^{LL} = \frac{(k - k_R)^2}{2m_R} + \Delta_R,
\end{equation}
where $m_R$ is an effective mass for the magnetorotons near the
minimum.  Here we have used the magnetoroton spectrum calculated in
the lowest Landau level.  Rappe, Zhu, and Louie \cite{rappe93} have
very recently calculated the magnetoroton spectrum as a function of
$r_s$ for $\nu = 1/3$, and find that at $r_s = 20$, the magnetoroton
spectrum has fallen only about 10\% below the lowest Landau level value.
At $\nu = $ 1/7 and 1/9 the amount of Landau level mixing found in
\cite{price93} is more than an order of magnitude less than that
found at $\nu = 1/3$, so using the lowest Landau level spectrum is an
excellent approximation.  Assuming that only the modes in the
vicinity of the minimum contribute, the free energy per particle of
the liquid is
\begin{equation}
F^{LL} = E^{LL}(r_s) - \left(2 \pi m_R \right)^{1/2} \frac{k_R
\ell^2}{\nu} T^{3/2} e^{-\Delta_R/T},
\label{free_liquid}
\end{equation}
which goes exponentially at small $T \leq \Delta_R$.

Because the excitations of the liquid display a gap, at very low
temperatures $T \ll \Delta_R$ even the lowest lying modes at the
minimum $k_R$ remain unoccupied, while the low-lying modes of the
solid, which have no gap, begin to fill immediately.  The free energy
of the solid then falls as a power of $T$, while the free energy of
the liquid remains nearly constant.  When the temperature begins to approach
the roton gap energy, however, the free energy of the liquid begins
to fall exponentially as the states become occupied.  Because the
density of states at the roton gap energy is very large, this
exponential rise is very rapid and the liquid free energy quickly
falls below that of the solid.

Figure \ref{energy_diff7} shows the difference in free energy $F^{WS}
- F^{LL}$ as a function of $T$ at $\nu = 1/7$ and 1/9.  The density $r_s =
2.3$ was chosen to match the density of one of the luminescence
experiments \cite{buhmann90}.  The difference in ground-state
energies favors the solid at zero temperature, and as the temperature
begins to rise, the solid is favored slightly more since the
magnetophonon modes are beginning to be occupied while the
magnetoroton modes are not.  Once the temperature becomes some
substantial fraction of the magnetoroton gap, however, the
exponential character of the free energy of the liquid begins to
assert itself as the magnetoroton modes become available, and the
liquid free energy rapidly drops below that of the solid.  The slope
at which the curves cross zero shows that the magnitude of the roton
gap $\Delta_R$ has much more effect on the melting temperature than
does any difference in ground-state energies $\Delta E = E^{WS} -
E^{LL}$, unless $\Delta E$ is very small.  Otherwise a change in
$\Delta E$, which shifts the curves up or down on the energy axis,
changes the melting temperature by only a small amount.

The phase boundaries for $\nu = $ 1/7 and 1/9 are shown in Figure
\ref{bound7}.  The relationship of the
melting temperature, shown as the solid lines, to the size of the
magnetoroton gap, shown as the dash-dotted lines, is clearly visible.
Melting, as a rule, will occur roughly at some constant fraction of
the magnetoroton gap.  The dashed lines in Figure \ref{bound7} show the
predicted classical Kosterlitz-Thouless melting
temperature.  The melting temperatures we find are roughly comparable
to the Kosterlitz-Thouless melting temperature, although there is no
theoretical reason to expect them to be closely related.

Disorder in the sample will have varying effects on the ground-state
and temperature-dependent parts of (\ref{free_liquid}) and
(\ref{free_solid}).  Impurities will cause the background charge in
the sample to become slightly non-uniform, and the Wigner solid will
adjust by compressing somewhat in areas of high background charge and
expanding somewhat in areas of low background charge.  The Laughlin
liquid, since it is incompressible, to lowest order cannot do this,
and the difference in ground-state energies $\Delta E = E^{WS} - E^{LL}$ will
change in
favor of the solid at very low temperatures.  In \cite{price93}, the
shift in energy due to impurities is estimated roughly as
\begin{equation}
E_{imp} \approx \frac{1}{2} m v_t^2 \left( \frac{a}{\xi} \right)^2
 = \frac{0.138}{r_s} \left( \frac{a}{\xi} \right)^2
\label{imp}
\end{equation}
where $v_t = (0.138e^2/ma)^{1/2}$ is the transverse
sound velocity in the absence
of the magnetic field and
$\xi$ is the correlation length for the distorted WS.  If we assume
$\xi = 5a$,
we find that the shift in ground-state energy is about $-0.0055/r_s$,
very significant at $\nu = 1/7$ and only slightly less so at $\nu =
1/9$.

On the other hand, the magnetoroton gap of the liquid is
significantly reduced by disorder \cite{boebinger85,macdonald86}.  At
$\nu = 1/5$, the measured gap (presumably the quasielectron-quasihole gap) is
1.1 K, while the single-mode
approximation of \cite{girvin86} which we have used gives $\Delta_R =
5.6$ K for the sample of \cite{jiang90}.  The luminescence
measurements give gaps on
the order of, but smaller than, 0.4 K at $\nu
= 1/7$, and 0.25 K at $\nu = 1/9$ for a sample
with $r_s = 2.3$.  The presence of disorder will open a small gap in
the magnetophonon spectrum, but this gap will have a negligible
effect on the free energy of the solid, since it is centered at the
origin of the Brillioun zone, where the density of states is small.

The result of moving the ground-state energy of the solid down by
$-0.0055/r_s$ and using $\Delta_R$ = 0.4 K at $\nu = 1/7$ and $\Delta_R$ = 0.25
K at $\nu = 1/9$, reduced from the corresponding theoretical results, is shown
in Figure \ref{energy_diff7}.  In spite of the shift in
ground-state energy favoring the solid at low temperatures, the
exponential drop in the free energy of the liquid near the
magnetoroton gap temperature moves the melting temperature down to
about 400 mK for both $\nu = 1/7$ and 1/9.  Because the measured gaps are
strictly speaking not the magnetoroton gap, but either the
quasielectron-quasihole gap or the magnetoroton mode at small $k$, the actual
gap $\Delta_R$ will be significantly lower than the measurements given above,
and the melting temperature will be proportionately lower as well.  Of course,
we do not know
the precise amount of disorder in the samples, but our calculation
shows that the Wigner solid may melt at a temperature equivalent or
slightly below those of  the luminescence experiments, while
remaining a solid at the lower temperatures of the transport
experiments.

With this melting in mind, we would like to propose a modification, shown in
Figure \ref{schematic}, to the phase diagram given in \cite{buhmann91}.  The
phase boundary at $\nu = 1/7$ and 1/9 no longer extends down to zero
temperature, as the previous authors proposed.  The ground state remains the
Wigner solid at these filling factors, but as the temperature is raised the
solid at $\nu = 1/7$ and 1/9 quickly gives way to a fractional quantum Hall
state.  Our
finite temperature
phase transition thus provides a possible explanation
for both the relatively high temperature magneto-optical
results \cite{buhmann90,buhmann91},
and the lower temperature
magneto-transport results \cite{jiang90,goldman88}.

Work at Berkeley was supported by NSF Grant No. DMR91-20269
and by the Director, Office of Energy Research,
Office of Basic Energy Sciences,
Materials Sciences Division of the U.S.
Department of Energy under contract No. DE-AC03-76SF00098.
CRAY computer time for work at Berkeley
was provided by the NSF at the San Diego
Supercomputing Center and by the Department of Energy.

\newpage
\figure{The difference in free energy between the Wigner solid and
the Laughlin liquid at $\nu = 1/7$ and 1/9, with $r_s = 2.3$.  The solid line
is computed using the theoretical value for the gap $\Delta_R$,
assuming no disorder, and the dashed line includes an estimate of
disorder-induced lowering of the Wigner solid ground-state energy and
uses $\Delta_R = $ 0.4 K at $\nu = 1/7$ and $\Delta_R = $ 0.25 K at $\nu =
1/9$.  The free energy is given in units of $(e^2/\epsilon)/2a_B$.
\label{energy_diff7}}

\figure{The phase boundary between Wigner solid and Laughlin liquid
(solid line) at $\nu = 1/7$ and 1/9.  The dashed line is the classical
Kosterlitz-Thouless melting temperature and the dash-dotted line is
the theoretical gap $\Delta_R$.  Disorder effects are not included in
this diagram.
\label{bound7}}

\figure{A qualitative phase diagram at finite but constant $r_s$.  The area
marked ``LL'' is the liquid region, and the areas marked ``WS'' are the solid
regions.  Temperature is given in arbitrary units.
\label{schematic}}

\end{document}